\documentstyle[11pt,newpasp,psfig,natbib]{article}

\def\ifundefined#1{\expandafter\ifx\csname#1\endcsname\relax}

\newif\ifpdf
\ifx\pdfoutput\undefined
   \pdffalse      
\else
   \pdfoutput=1   
   \pdftrue
\fi

\def\la{\mathrel{\hbox{\rlap{\hbox{\lower4pt\hbox{$\sim$}}}\hbox{$<$}}}}
\def\ga{\mathrel{\hbox{\rlap{\hbox{\lower4pt\hbox{$\sim$}}}\hbox{$>$}}}}

\newcommand{\be}{\begin{eqnarray}}
\newcommand{\ee}{\end{eqnarray}}

\ifundefined{ensuremath}\def\ensuremath#1{\relax\ifmmode{#1}}
\else${#1}$\fi\else\relax\fi
\ifundefined{nuc}\def\nuc#1#2{\relax\ifmmode{}^{#1}{\protect\text{#2}}
\else${}^{#1}$#2\fi}\else\relax\fi

\def\tstd{\ensuremath{\tau_{\textrm{std}}}}

\def\alog#1{\times 10^{#1}}
\newcommand{\phx}{\texttt{PHOENIX}}

\def\la{\mathrel{\hbox{\rlap{\hbox{\lower4pt\hbox{$\sim$}}}\hbox{$<$}}}}
\def\ga{\mathrel{\hbox{\rlap{\hbox{\lower4pt\hbox{$\sim$}}}\hbox{$>$}}}}

\renewcommand{\be}{\begin{eqnarray}}
\renewcommand{\ee}{\end{eqnarray}}

\def\SP2{{\tt IBM SP2}}

\def\PC2{{\tt PC$^2$}}

\def\inu{\ifmmode{I_{\nu}}\else{\hbox{$I_{\nu}$} }\fi}
\def\snu{\ifmmode{S_{\nu}}\else{\hbox{$S_{\nu}$} }\fi}
\def\jnu{\ifmmode{J_{\nu}}\else{\hbox{$J_{\nu}$} }\fi}

\def\fep{\ifmmode{{\rm Fe II}}\else\hbox{Fe~II }\fi}

  at 12.0 true pt 

\def\phx{{\tt PHOENIX}}

\def\phx{{\tt PHOENIX}}

\def\g{\gamma}
\def\b{\beta}
\def\m{\mu}

\def\l{\lambda}
\def\L{\Lambda}

\def\pder#1#2{{\partial #1 \over \partial #2}}
\def\div#1#2{{#1\over #2}}
\def\rout{\ifmmode{r_{\rm out}}\else\hbox{$r_{\rm out}$}\fi}
\def\tmax{\ifmmode{\tau_{\rm max}}\else\hbox{$\tau_{\rm max}$}\fi}
\def\tstd{\ifmmode{\tau_{\rm std}}\else\hbox{$\tau_{\rm std}$}\fi}
\def\vmax{\ifmmode{v_{\rm max}}\else\hbox{$v_{\rm max}$}\fi}
\def\muE{\ifmmode{\mu_{\rm E}}\else\hbox{$\mu_{\rm E}$}\fi} 
\def\pE{\ifmmode{p_{\rm E}}\else\hbox{$p_{\rm E}$}\fi} 
\def\bmax{\ifmmode{\b_{\rm max}}\else\hbox{$\b_{\rm max}$}\fi}

\def\alog#1{\times 10^{#1}}

\def\rout{\hbox{$r_{\rm out}$} }

\def\chistd{\ifmmode{\chi_{\rm std}}\else\hbox{$\chi_{\rm std}$}\fi}

\def\lstar{\ifmmode{\Lambda^*}\else\hbox{$\Lambda^*$}\fi} 
\def\Rop{\ifmmode{[R_{ij}]}\else\hbox{$[R_{ij}]$}\fi}

\def\Rji{\ifmmode{[R_{ji}]}\else\hbox{$[R_{ji}]$}\fi}
\def\Rstar{\ifmmode{[R_{ij}^*]}\else\hbox{$[R_{ij}^*]$}\fi}
\def\Rijstar{\Rstar}
\def\Rjistar{\ifmmode{[R_{ji}^*]}\else\hbox{$[R_{ji}^*]$}\fi}
\def\DRji{\ifmmode{[\Delta R_{ji}]}\else\hbox{$[\Delta R_{ji}]$}\fi}
\def\DRij{\ifmmode{[\Delta R_{ij}]}\else\hbox{$[\Delta R_{ij}]$}\fi}

\def\ns{\ifmmode{N_{\rm s}}          
        \else\hbox{$N_{\rm s}$}\fi}

\def\mat#1{{\bf #1}}     
\def\vek#1{{#1}}         

\newcount\eqcount
\def
  \nummer{
    \global\advance\eqcount by 1
    (\the\eqcount)
  }

\def
  \numadv{
    \global\advance\eqcount by 1
  }

\def
   \numout#1{
     (\the\eqcount #1)
  }

\def\ivek#1#2{\ifmmode{\vek{I}^{#1}_{#2}}
        \else\hbox{$\vek{I}^{#1}_{#2}$}\fi}

\def\tmat#1#2{\ifmmode{\mat{t}^{#1}_{#2}}
        \else\hbox{$\mat{t}^{#1}_{#2}$}\fi}
\def\rmat#1#2{\ifmmode{\mat{r}^{#1}_{#2}}
        \else\hbox{$\mat{r}^{#1}_{#2}$}\fi}
\def\bvek#1#2{\ifmmode{\beta^{#1}_{#2}}
        \else\hbox{$\beta^{#1}_{#2}$}\fi}

\def\lp{\ifmmode{\lambda^+_\tau}           
        \else\hbox{$\lambda^+_\tau$}\fi}
\def\lm{\ifmmode\lambda^-_\tau             
        \else\hbox{$\lambda^-_\tau$}\fi}

\begin{document}

\title{Highlights of Stellar Modeling with PHOENIX}

\author{E. Baron\altaffilmark{1}, Peter
H.~Hauschildt\altaffilmark{2,3}, F.~Allard\altaffilmark{4}, Eric
J.~Lentz\altaffilmark{2}, 
J.~Aufdenberg\altaffilmark{5}, 
A.~Schweitzer\altaffilmark{2,3}, and T.~Barman\altaffilmark{2,6}}



\altaffiltext{1}{Dept. of Physics and Astronomy, University of
Oklahoma, 440 W. Brooks, Rm.~131, Norman, OK 73019 USA}
\altaffiltext{2}{Dept. of Physics \& Astronomy and Center for
Simulational Physics, University of Georgia, Athens, GA 30602-2451
USA}
\altaffiltext{3}{Present Address: Hamburger Sternwarte, Gojenbergsweg 112,
21029 Hamburg, Germany}
\altaffiltext{4}{CRAL,
Ecole Normale Superieure,
46 Allee d'Italie, Lyon,
69364 France, Cedex 07}
\altaffiltext{5}{Harvard-Smithsonian   
Center for Astrophysics,      Mail Stop 15,           60 Garden Street      
Cambridge, MA 02138 USA}
\altaffiltext{6}{Present Address: Dept. of Physics, Wichita State University,
Wichita, KS 67260-0032 USA}


\setcounter{footnote}{6}


\begin{abstract}
We briefly describe the current version of the \phx\ code. We then
present some illustrative results from the modeling of Type Ia
and Type II supernovae, hot stars, and irradiated giant planets. Good
fits to observations can be obtained, when account is taken for
spherically symmetric, line-blanketed, static or expanding
atmospheres. 
\end{abstract}




%
%
%

\section{Introduction}

\phx\ \citep[see][and references therein]{hbjcam99} is a generalized,
stellar model atmosphere code for treating both static and moving
atmospheres. The goal of \phx\ is to be both as general as possible so
that essentially all astrophysical objects can be modeled with a
single code, and to make as few approximations as possible. 
Approximations are inevitable (particularly in atomic data where
laboratory values for most quantities are unknown); however, the
agreement of synthetic spectra with observations across a broad class
of astrophysical objects is a very good consistency check. 
We have modeled 
Planets/BDs \citep{barmanna102,cond-dusty,lhires,latel}, Cool
Stars \citep{nextgen199,nextgen299}, Hot Stars \citep[$\beta$CMa,
$\epsilon$CMa, Deneb][]{aufdenecm98,aufdencma99,aufden02},
$\alpha$-Lyra, Novae \citep{phhnovetal97,novand86pap}, and all types of
superovae   \citep[SNe Ia, Ib/c, IIP,
IIb][]{b93j4,l94d01,b94i2,mitchetal87a02}.

\section{The \phx\ Code}

\phx\ solves the radiative transfer problem by using the
short-characteristic method \citep{OAB,ok87} to obtain the formal
solution of the special relativistic, spherically symmetric radiative
transfer equation (SSRTE) along its characteristic rays. The
scattering problem is solved via the use of  a
band-diagonal approximation to the discretized $\Lambda$-operator
\citep{phhs392,ok87,hsb94} as our choice of the approximate
$\Lambda$-operator.  This method can be implemented very efficiently
to obtain an accurate solution of the SSRTE for continuum and line
transfer problems using only modest amounts of computer resources.

We emphasize that \phx\ solves the radiative transfer problem
including a full treatment of special
relativistic radiative transfer in 
spherical geometry for all lines and continua. In addition we enforce
the generalized condition of 
radiative equilibrium in the Lagrangian frame, including
all velocity terms and deposition of energy from radiative decay or
from external irradiation. 

We also include a full non-LTE treatment of most ions, using model
atoms constructed from the data of \citet{kurucz93,cdrom22,cdrom23}
and/or from the CHIANTI database {\tt
http://wwwsolar.nrl.navy.mil/chianti.html}. The code uses Fortran-95
data structures to access the different databases and model atoms from
either or both databases can be selected at exectution time.

Absorption and emission is treated assuming complete redistribution
and detailed 
depth-dependent profiles for the lines. Fluorescence effects are
included in the NLTE treatment.
The equation of state used includes up to 26
ionization stages of 40 elements as well as up to 206 molecules.

The atomic data is constructed from
all relevant b-f and f-f transitions \citep{mathisen84,VY95,SYMP94}
as well as collisional rates obtained from laboratory measurements
(where available), the Opacity Project, the Iron Project, Van
Regemorter's formula \citep{vr62}, and the semi-empirical formula from
\citet{allen_aq}.

In addition to NLTE lines, lines treated in LTE and are
selected dynamically from the $42\alog{6}$ list of \citet{kurucz93}. A
typical calculation includes  about $2\alog{6}$ lines that are treated
as background opacity \citep{hbjcam99}. Molecular opacities (about
$7\alog{8}$ lines) are
discussed in \citet{cond-dusty} and molecular NLTE is discussed in
\citet{SHB00}.

\section{The equation of radiative transfer}

The equation of transfer in spherical coordinates in the co-moving
frame can be written \citep{found84}:
\vskip 0.5 true cm
$$
    e \pder{I}{r}
      + \pder{}{\m} \left( fI \right)
      + g \pder{}{\l} \left( \l I \right)
      + h I
    = \eta - \chi I 
$$
with 
\begin{eqnarray*}
   e(r,\m) &= &\g (\m+\b) \\
   f(r,\m) &= &\g (1-\m^2)
              \left[\div{1+\b\m}{r}-\g^2(\m+\b)\pder{\b}{r} \right]
                         \\
   g(r,\m) &= &\g \left[\div{\b(1-\m^2)}{r}
              +\g^2\m(\m+\b)\pder{\b}{r} \right]
                                     \cr
   h(r,\m) &= &\g \left[\div{\b(1-\m^2)}{r}
              +\g^2(1+\m^2+2\b\m)\pder{\b}{r} \right]
                         \\
\end{eqnarray*}
where {$I(r,\mu,\lambda)$} is the specific intensity scaled by $r^2$,
{$r$} is the radial coordinate,
$\mu=\cos\theta$ is the cosine of the direction angle,
{$v$} is the velocity, $\b = v/c$, $\g^2 = 1/(1-\b^2)$,
{$\chi(r,\lambda)$} is the total 
extinction coefficient, $\chi=\kappa+\sigma_e+\kappa_l\varphi(\lambda)$,
and
{$\eta(r,\lambda)$} is the  emissivity.

\subsection{Numerical solution}

The details of the numerical solution are discussed in detail in
\citet{hbjcam99} so we only sketch the basic idea here.
We discretize the $\partial/\partial \lambda$ term and
treat the boundary value problem for each wavelength individually,
using an Operator splitting (OS) method.
The steps are: solve along characteristics of the RTE,
using 
the piecewise parabolic ansatz to calculate $I$ for given $J$,
and iterate to self-consistent solution for $J$ using a band diagonal
accelerated lambda operator. 
The eigenvalues of amplification matrix are close to unity 
and the  use of operator splitting reduces the  eigenvalues of the
amplification 
matrix.

\section{Statistical Equilibrium Equations}

The radiative transfer equation depends on the opacities and
emissivity, which in turn depend on the level populations (which in
turn depend on the radiation field). Thus, we must iterate the
statistical equilibrium equations, simultaneously with the radiative
transfer equation \citep[the solution of the generalized radiative
equilibrium condition is discussed in][]{phhtc03}. 

The rate equations are given by \citep[see ][]{found84}
\begin{eqnarray*}
\lefteqn{\sum_{j<i} n_j \left(R_{ji}+C_{ji}\right)} \\
  & &  { -n_i\left\{\sum_{j<i}
		\left(R_{ij}+C_{ij}\right)
   +\sum_{j>i} \left(\div{n_j^{*}}{n_i^{*}}\right)
		\left(R_{ij}+C_{ij}\right)\right\} }   \\
 & &  {    +\sum_{j>i} n_j
     \left(\div{n_j^{*}}{n_i^{*}}\right)\left(R_{ji}+C_{ji}\right)
     =  0.} \\
\end{eqnarray*}

\subsection{Solution of the Rate Equations}

Note that line and continuum scattering
prevents the use of the $\L$-iteration for the 
solution of the rate equations.
Therefore, we use the Operator Splitting method (pre-conditioning)
to 
define a ``rate operator'' in analogy to the $\L$-operator \citep{phhcas93}:
$$
 R_{ij} = [R_{ij}][n].
$$
We also define an ``approximate rate operator'' $\Rijstar$ and
write the iteration scheme in the form: 
$$
     R_{ij} = \Rijstar [n_{\rm new}] + \left([R_{ij}]-\Rijstar\right) [n_{\rm old}].
$$

Inserting the above expression for $R_{ij}$ into the statistical
equilibrium equations and linearizing we obtain,
\begin{eqnarray*}
\lefteqn{\sum_{j<i} n_{j,\rm old} [R_{ji}^{*}][n_{\rm new}]} \\
 & -& n_{i,\rm old}\left\{\sum_{j<i}
		[R_{ij}^{*}][n_{\rm new}]
	     +\sum_{j>i} \left(\div{n_j^{*}}{n_i^{*}}\right)
		 [R_{ij}^{*}][n_{\rm new}]\right\}     \\
 & +&\sum_{j>i} n_{j,\rm old}
     \left(\div{n_j^{*}}{n_i^{*}}\right)[R_{ji}^{*}][n_{\rm new}]\\
 &    +&\sum_{j<i} n_{j,\rm new} \left(\DRji[n_{\rm old}]+C_{ji}\right) \\
 & -&n_{i,\rm new}\left\{\sum_{j<i}
		\left(\DRij[ n_{\rm old}]+C_{ij}\right)\right. \\
 & &\qquad\quad\quad \left.+\sum_{j>i} \left(\div{n_j^{*}}{n_i^{*}}\right)
		 \left(\DRij[n_{\rm old}]+C_{ij}\right)\right\}      \\
   &+&\sum_{j>i} n_{j,\rm new}
     \left(\div{n_j^{*}}{n_i^{*}}\right)\left(\DRji[n_{\rm old}]+C_{ji}\right)
     = 0. \\
\end{eqnarray*}

\section{The Computational Problem}

The size of the computational problem is determined by:
\begin{itemize}
\item {\bf input data size}
\begin{itemize}
\item number of atomic/ionic spectral lines: $\approx 42\alog{6}$ $\to 0.6\,$GB

\item number of diatomic molecular lines: $\approx 35\alog{6}$ $\to 0.5\,$GB

\item number of hot water vapor lines: 
\item number of hot water vapor lines:
\begin{item}
\itemsep=0pt
\item before 1994: $\approx 0.035\alog{6}$
\item 1994: $+\approx 6\alog{6}$ $\to 0.1\,$GB
\item 1997: $+\approx 330\alog{6}$ $\to 5\,$GB
\item 2001: $+\approx 100\alog{6}$ $\to 1.5\,$GB
\end{item}
\item total molecular lines (May 2001): $\approx 700\alog{6}$ $\to 10\,$GB
\end{itemize}

{\em all} lines need to be accessed in a {\em line selection}
procedure which
dynamically  creates sub-lists that can be as large as 
the original list, but are generally much smaller. This
poses a significant data handling problem, which we have surmounted
via the use of both memory and disk caches that allows us 
to trade (at run time) available memory for I/O bandwidth if required on a
particular machine or architecture.

\item {\bf memory/IO requirements}

\begin{itemize}

\item number of individual energy levels: $\approx 10,000$ $\to$ $\approx 10\,$MB

\item number of individual NLTE transitions:\\
$\approx 100,000$ $\to$ $\approx 150\,$MB

\item EOS data storage $\approx 40\,$MB

\item auxiliary storage $\approx 50\,$MB

\item $\to$ total memory requirement $\ge 200\,$MB

\item number of individual energy levels and transitions has increased
dramatically $\to$ memory requirements $>0.5\,$GB with the inclusion
of more molecular species and the Chianti database.

\end{itemize}

\item {\bf (serial) CPU time}

\begin{itemize}

\item small for each individual point on the wavelength grid:
$\approx 10\ldots 100\,$msec

\item number of wavelength points for radiative transfer:
30,000-500,000 (can be $>10^6$)

\item $\to$ $\approx 50,000$~sec to ``sweep'' once through all wavelength points
\item typically, $\approx 10$ iterations (sweeps) are required to 
obtain an equilibrium model

\item $\to \approx 6\,$ CPU {\em days}

\item there are, literally, 1000's of models in a typical grid \ldots

The solution is parallel computing, which we have implemented in a
MIMD model \citep{hbapara97,bhpar298,hlb01,bpar03}. This dramatically
reduces wallclock time per model and makes very large scale models feasible.
\end{itemize}

\end{itemize}

These numbers above are for models with 50 layers, with modern
supercomputers many models use 100 or more layers and the scaling goes
approximately as the cube of the number of layers.

\section{Applications}

\begin{figure}[ht]
\begin{center}
\leavevmode
\psfig{file=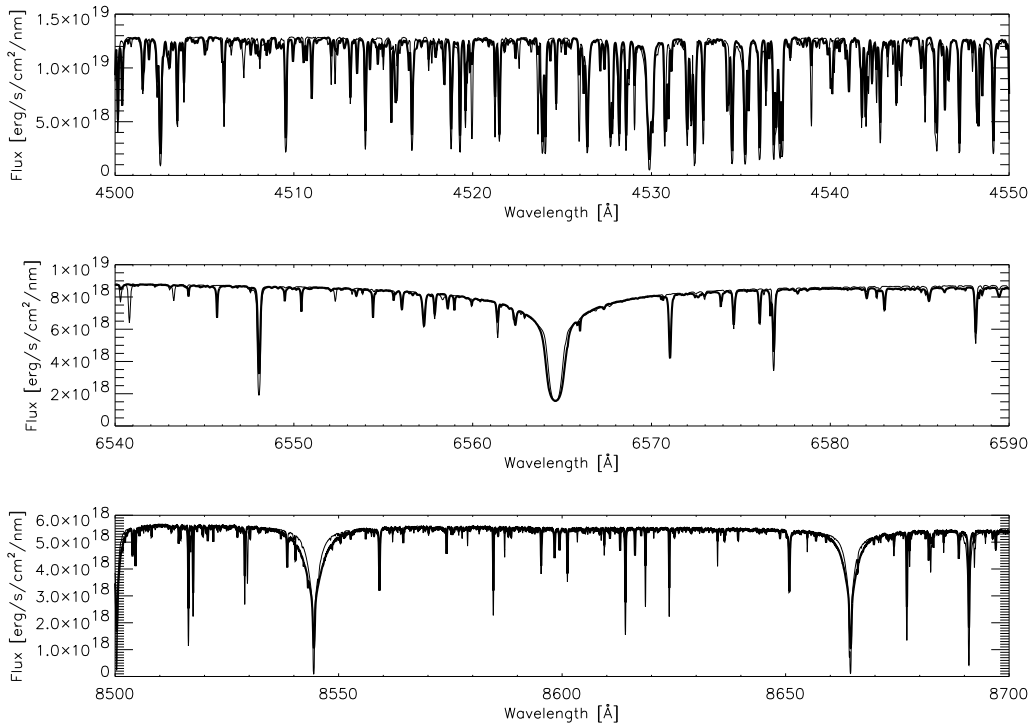,height=10cm,angle=90}
\caption{The synthetic model spectrum of the Sun (thin line) compared
to observations. 
\label{fig:thesun}}
\end{center}
\end{figure}

\begin{figure}[ht]
\begin{center}
\leavevmode
\psfig{file=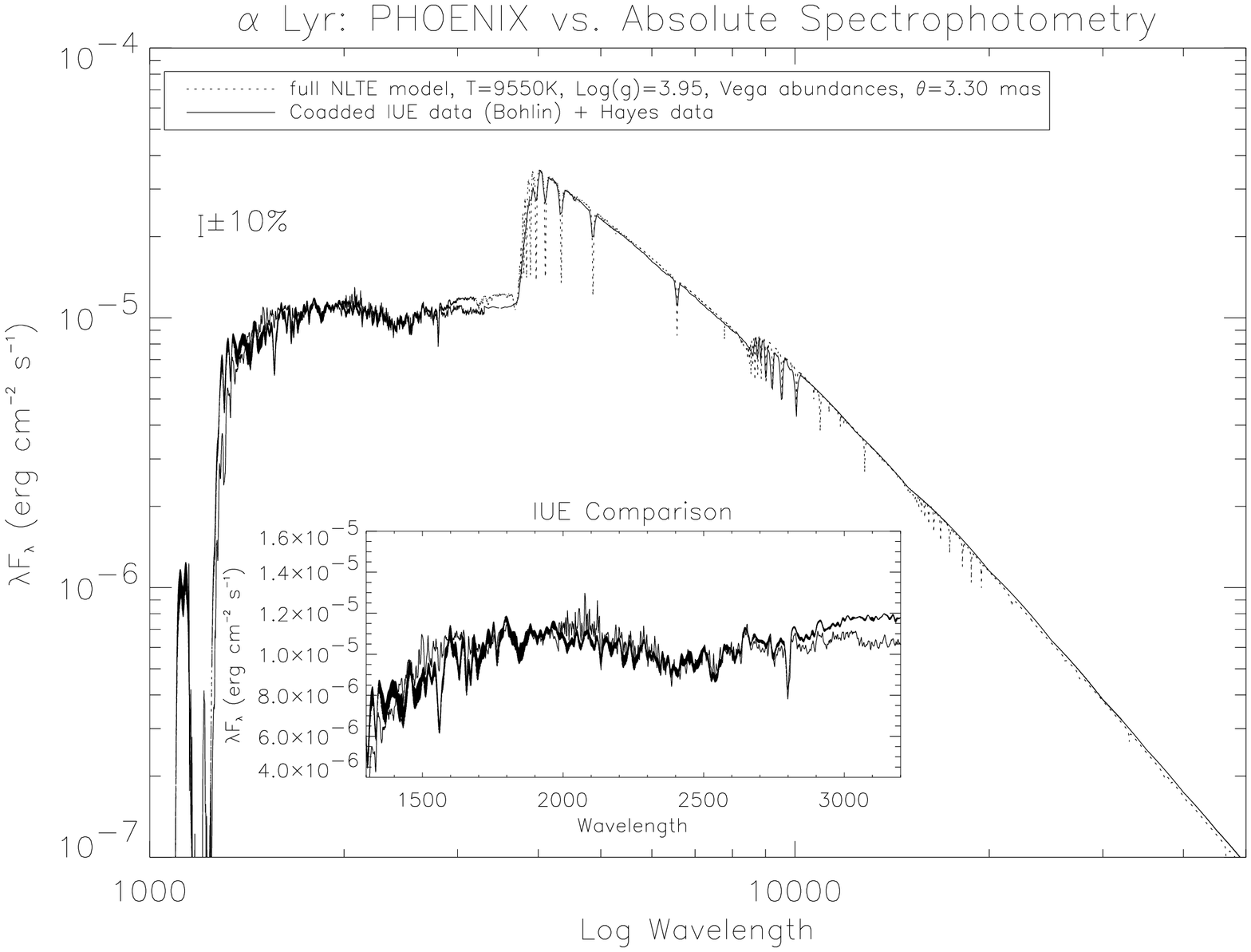,height=10cm,angle=90}
\caption{The synthetic model spectrum of Vega compared to observations
\citep{vegaopt85,vegaiue90}. \label{fig:vega}}
\end{center}
\end{figure}

We show a few illustrative applications of
\phx. Figure~\ref{fig:thesun} displays the observed and synthetic
spectrum of the Sun and the agreement is quite
good. Figure~\ref{fig:vega} shows the observed and synthetic spectra
of Vega, with excellent agreement in the UV and only a few features
that are too strong in the optical compared to observations.

%
%
%

%

\begin{figure}[ht]
\begin{center}
\leavevmode
\psfig{file=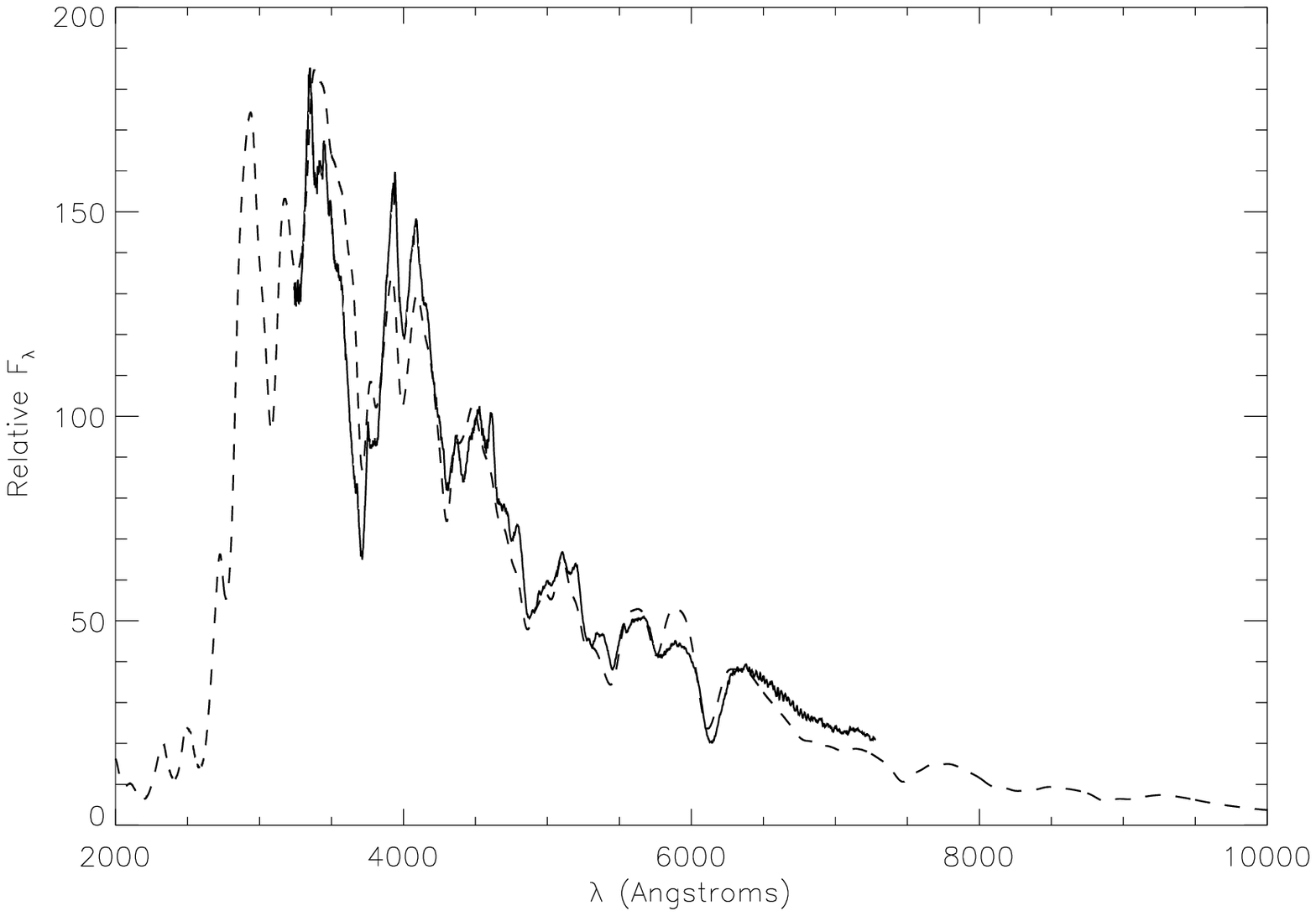,height=8cm}
\caption{SN 1994D on 15~March 1994 \protect\citep[solid line,][]{filipasi97}
and W7 best fit synthetic spectrum for
day 14 after explosion (dashed line), from \protect\citet{l94d01}.\label{fig:d14}}
\end{center}
\end{figure}

\begin{figure}[ht] 
\begin{center}
\leavevmode
\psfig{file=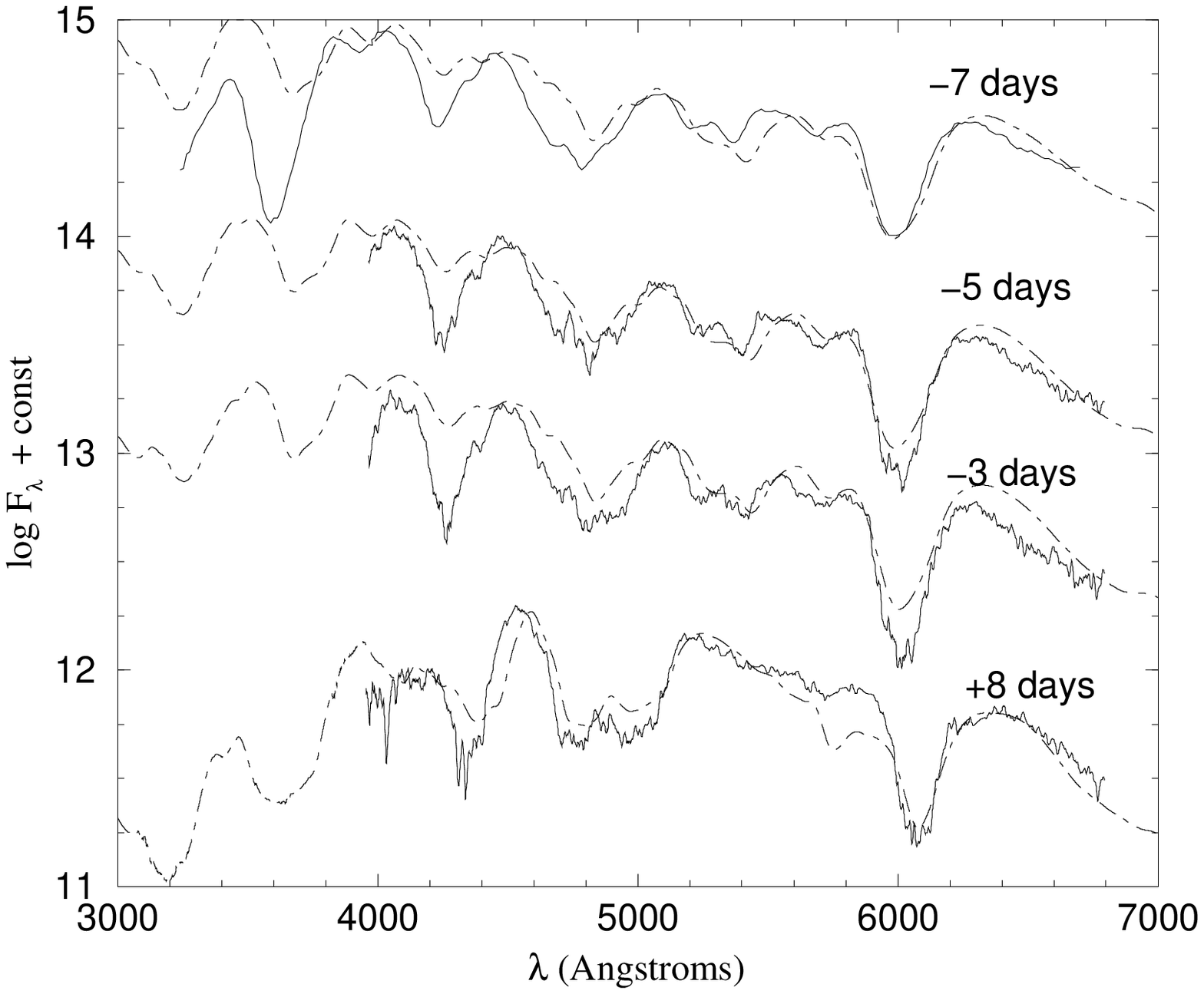,height=7cm,angle=0}
\caption{Synthetic spectra
for the delayed detonation model 21c~(dot-dashed lines) plotted
against observed spectra for 
SN~1984A~(solid lines) from \protect\citet{l84a01}.\label{fig:dd21c}} 
\end{center}
\end{figure}

Turning from stars to supernovae, we note that supernovae are
fundamentally different from stars, since in stars spectra only probe
to the photosphere, whereas a time series of supernovae spectra allow
us to ``peel the onion'' since geometrical dilution due to expansion
of the ejecta causes the
photosphere to receed in mass with time. Figure~\ref{fig:d14} shows
the synthetic spectrum of the W7 deflagration model \citep{nomw7}
compared to the 
observations of SN~1994D at 6 days prior to maximum light
\citep{l94d01}. Figure~\ref{fig:dd21c} shows that delayed detonation
models well reproduce the spectra of SN~1984A --- a ``fast'' SNe~Ia
\citep{l84a01,hathighv00}.

\begin{figure}[ht] 
\begin{center}
\leavevmode
\psfig{file=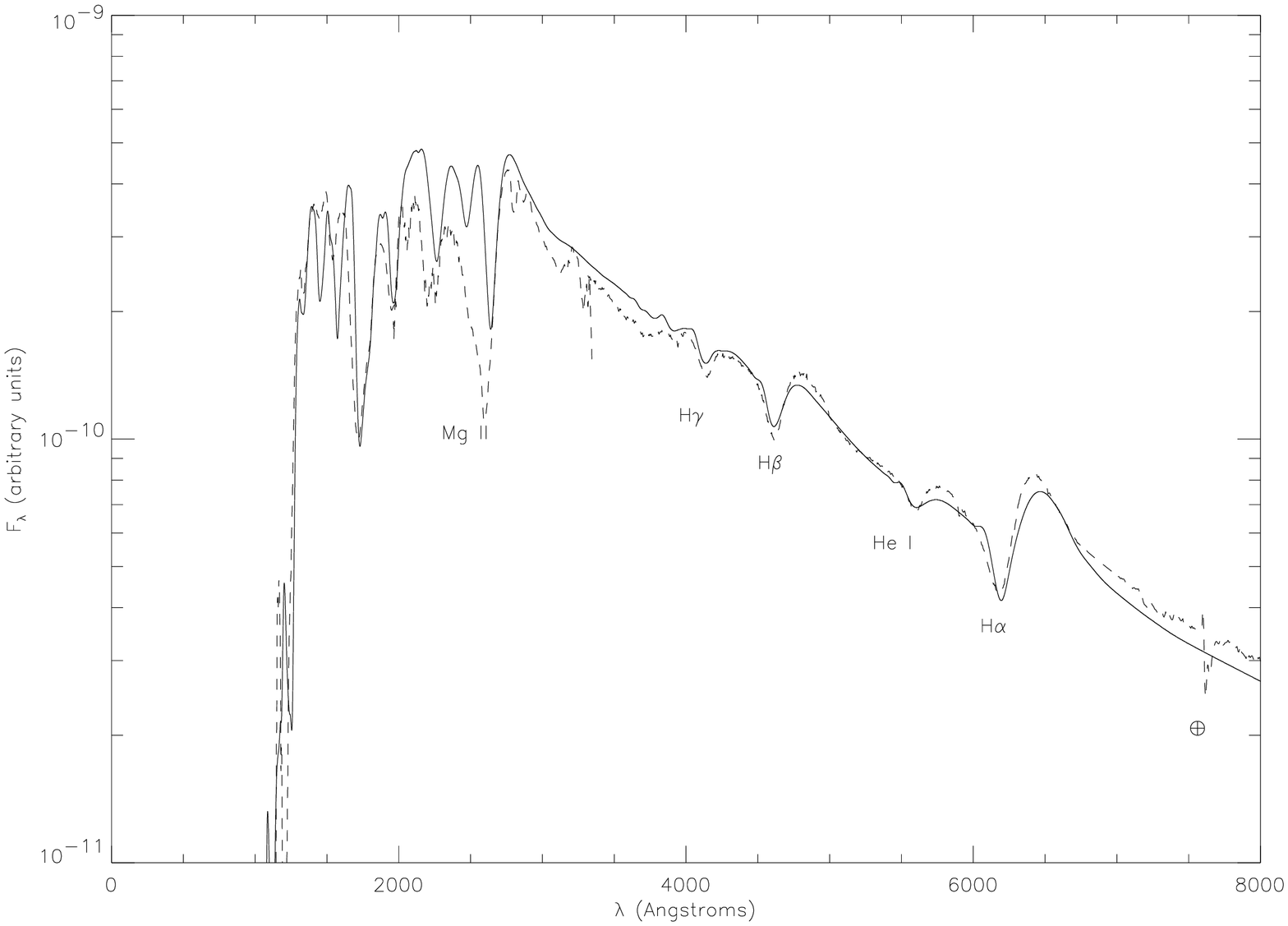,height=8cm,angle=90}
\caption{\label{fig:1opt} \phx\ model spectrum (solid line) for
day~1.36 \protect\citep{mitchetal87a02}. Important optical lines
include H$\alpha$ through H$\delta$, 
and He~I~$\lambda$5876.  The optical
spectra are taken from the CTIO archive \citep{ctio87a88} and all UV
spectra are from IUE \citep{punetal95}.} 
\end{center}
\end{figure}

\begin{figure}[ht] 
\begin{center}
\leavevmode
\psfig{file=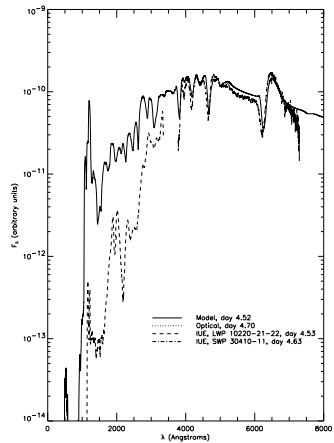,height=6cm,angle=0}
\caption{\label{fig:7nopt} \phx\ model spectrum for SN~1987A,
day~4.52, with gamma-ray 
deposition calculated assuming local deposition due to a constant nickel mass
fraction of $1.0 \times 10^{-3}$ everywhere in the envelope
\citep{mitchetal87a02}.} 
\end{center}
\end{figure}

\begin{figure}[ht] 
\begin{center}
\leavevmode
\psfig{file=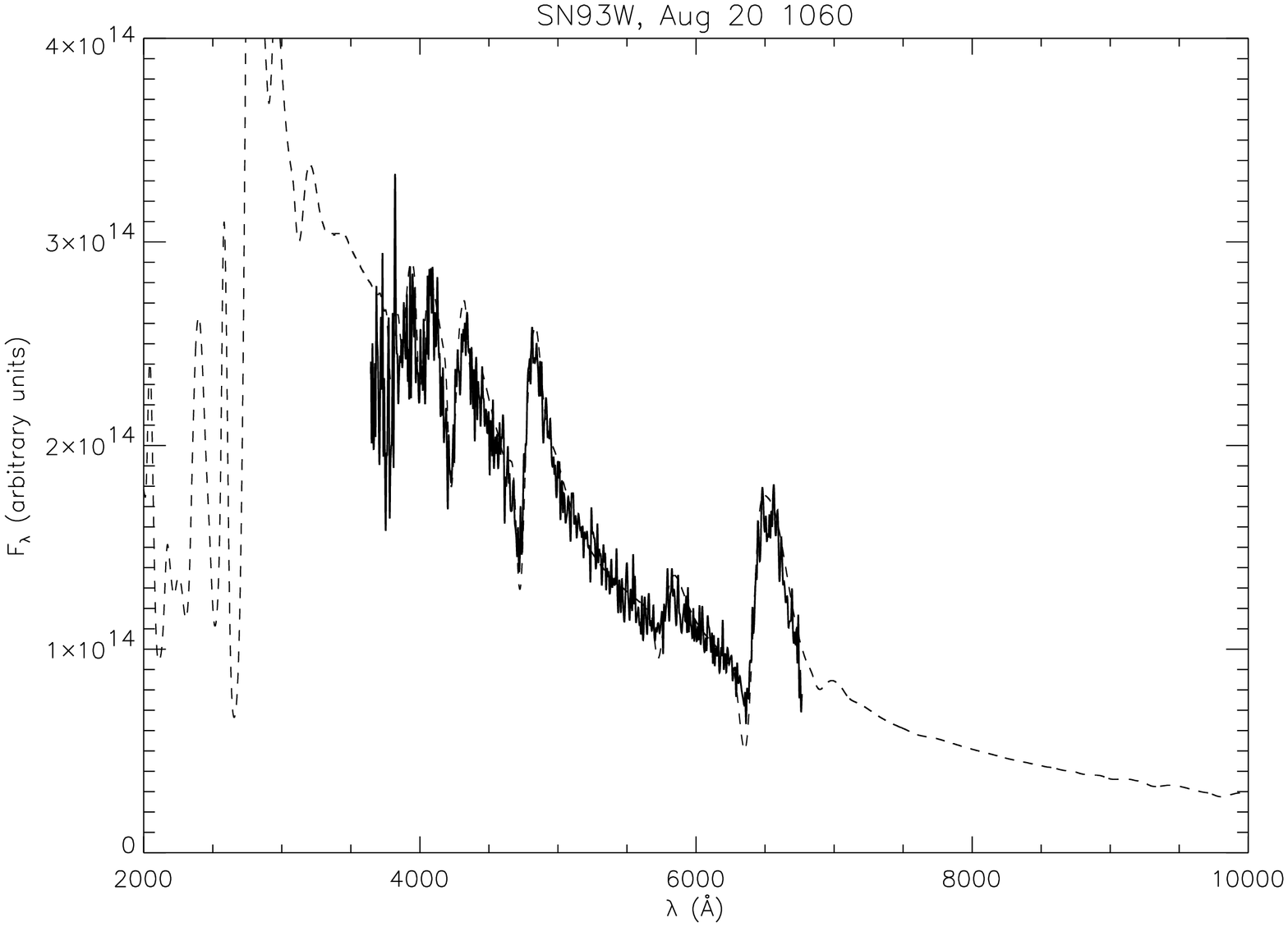,height=10cm,angle=90}
\caption{The observed spectrum of the Type IIP SN~1993W compared to the
synthetic spectrum. The agreement is excellent. \label{fig:93w}}
\end{center}
\end{figure}

We have also applied \phx\ to core collapse
supernovae. Figure~\ref{fig:1opt} shows a very good fit to the first
spectrum obtained of SN~1987A in both the UV and optical and
Figure~\ref{fig:7nopt} 
shows the \phx\ synthetic spectrum 
compared to SN~1987A at 4.5 days after the explosion. In order to
reproduce H$\alpha$, extra nickel mixing in the envelope was
required. It is evident that the simple nickel mixing parameterization
has destroyed the fit in the UV and clumping is probably required to
fit the entire spectrum. Figure~\ref{fig:93w} shows that excellent
fits to normal Type IIP spectra can be obtained at early times. These
fits allow us to determine the reddening, amount of H/He mixing,
amount of nickel mixing, and the progenitor metallicity.

%
%

\section{Conclusions}

We have described the basic design of the generalized model atmosphere
code \phx. We have shown that good, but not perfect agreement with a
variety of astrophysical objects can be obtained. Future work will
include  full time dependent calculations and full 3-D detailed NLTE
radiative transfer.


\acknowledgments
 We thank our many collaborators who have
contributed to the development of {\tt PHOENIX}, in particular 
Ian Short, David
Branch, Sumner Starrfield, and Steve Shore. 
This work was supported in part by NSF grant
AST-9720704, NASA ATP grant NAG 5-8425 and LTSA grant NAG 5-3619 to
the University of Georgia and by NASA grants
NAG5-3505, NAG5-12127, NSF grant AST-0204771, and an IBM SUR grant to
the University of Oklahoma. PHH was 
supported in part by the P\^ole Scientifique de Mod\'elisation
Num\'erique at ENS-Lyon. Some of the calculations presented here were
performed at the San Diego Supercomputer Center (SDSC), supported by
the NSF, and at the National Energy Research Supercomputer Center
(NERSC), supported by the U.S. DOE.  We thank both these institutions
for a generous allocation of computer time.

\clearpage



\end{document}